\documentclass[12pt,thmsa,a4paper,oneside,onecolumn]{article}
\usepackage{amssymb}

\usepackage[]{cite}

\input{tcilatex}
\begin{document}

\author{Yi-Hang Nie$^{1,2}$,Yan-Hong Jin$^{1,5}$, \and J.-Q.Liang$^{1,3}$%
,H.J.W.M\"{u}ller-Kirsten$^3$,D.K.Park$^{3,4}$,F.-C.Pu$^{5,6}${\small \ } \\
$^1${\small Institute of Theoretical Physics and Department of
Physics,Shanxi University,}\\
{\small Taiyuan,Shanxi 030006,\ People}$^{,}${\small s Republic of China}\\
$^2${\small Department of physics,Yanbei Normal Institute, Datong Shanxi
037000, }\\
{\small People}$^{,}${\small s Republic of China}\\
{\small \ \ }$^3${\small Department of Physics, University of\
Kaiserslautern D-67653 Kaiserslautern,Germany}\\
{\small \ \ }$^4${\small Department of Physics, Kyungnam University, Masan,
631-701, Korea}\\
{\small \ \ }$^5${\small Institute\ of\ Physics\ and\ Center for Condensed
Matter Physics\ ,}\\
{\small Chinese Academy of Sciences,} {\small Beijing 100080,People}$^{,}$%
{\small s Republic of China, \ }\\
$^6${\small \ Department of Physics, Guangzhou Normal College, Guangzhou
510400,}\\
{\small People}$^{,}${\small s Republic of China }}
\title{Macroscopic Quantum Phase Interference in\\
Antiferromagnetic Particles }
\date{{\small 20 Dec. 1999 }}
\maketitle

\begin{abstract}
\textrm{The tunnel splitting in biaxial antiferromagnetic particles is
studied with a magnetic field applied along the hard anisotropy axis. We
observe the oscillation of tunnel splitting as a function of the magnetic
field due to the quantum phase interference of two tunneling paths of
opposite windings. The oscillation is similar to the recent experimental
result with Fe}$_8$\textrm{\ molecular clusters.}
\end{abstract}

The macroscopic quantum phenomenon of magnetic particles at low temperature
has attracted considerable attention both theoretically and experimentally
in recent years\cite{E.M.Chud,B.Barbara,Proceeding}. The magnetization
vector in solids is traditionally viewed as a classical variable. The
quantum transition of the magnetization vector\textbf{\ M} between different
easy directions in a single domain ferromagnetic (FM) grain, in particular
the coherent tunneling between two degenerate orientations of the
magnetization called the macroscopic quantum coherence (MQC)\cite{A.Garg PRL}%
, has been studied extensively for its exotic characters far from that of
classical system. Quenching of MQC for half-integer spin is a fascinating
effect \cite{D.Loss,EMC DPD,N.V.P,J.Q.Liang}and can be used to test the
macroscopic quantum tunneling experimentally. The quenching of MQC in spin
particles is analysed with the help of the phase interference of spin
coherent state-paths which possess a phase with an obvious geometric meaning%
\cite{D.Loss}. Although the quenching of MQC has been interpreted as Kramers$%
^{\text{,}}$ degeneracy, the effect of geometric phase interference is far
richer than that. By investigating the quantum tunneling in biaxial
ferromagnetic particles with a magnetic field applied along the hard axis
Garg\cite{A.Garg} found a new quenching of tunneling splitting which is not
related to Kramers$^{,}$ degeneracy since the external field breaks the time
reversal symmetry. The Zeeman energy of the biaxial spin particle in the
external magnetic field results in additional topological phases of the
tunnel paths which lead to the quantum phase interference. The tunneling
splitting therefore oscillates with respect to the magnetic field.

According to a recent report\cite{W.Werns} the oscillation of tunneling
splitting was observed experimentally in molecular clusters Fe$_8$ which at
low temperature behavior like a nanomagnet, namely, a ferromegnetic
particle. A more detailed analysis of quantum phase interference with
instanton method in context of spin coherent-state-path-integrals has been
given recently \cite{S.P.Kou}. In the present letter we investigate the
similar effect of quantum phase interference in antiferromagnetic (AFM)
particles. Since the tunneling rate in AFM particles is much higher than
that in the FM particles of the same volume\cite{J.M.Duan}, the\ AFM
particles are expected to be a better candidate for the observation of MQP
than the FM particles\cite{J.M.Duan}. The AFM particle is usually described
by the N\'{e}el vector of the two collinear sublattices whose magnetizations
are coupled by strong exchange interaction. External magnetic field dose not
play a role since the net magnetic moment vanishes for idealized
sublattices. The quantum and classical transitions of the N\'{e}el vector in
antiferromagnet have been well studied\cite{H.Sima} in terms of the
idealized sublattice model. The temperature dependence of quantum tunneling
was also given for the same model\cite{J.Q.L Mu Nie} and the theoretical
result agrees with the experimental observation\cite{J.Tejada}. A biaxial
AFM particle with a small non-compensation of sublattices in the absence of
an external magnetic field was studied in Ref.\cite{E.M.C.in} where it was
shown that the uncompensated magnetic moment leads to a modification of the
oscillation frequency around the equilibrium orientations of the N\'{e}el
vector. In the present letter we demonstrate that the uncompensated magnetic
moment of a small biaxial AFM particle possesses a Zeeman energy in an
external magnetic field applied along hard axis and thus a topological phase
depending on the magnetic field is introduced similarly to the phase in
ferromagnetic particles\cite{A.Garg}. The quantum phase interference results
in the oscillation of tunneling splitting as a function of magnetic field
which may be regarded as a kind of the Aharonov-Bohm effect.

We consider in the following a biaxial AFM particle of two collinear FM
sublattices with a small non-compensation. Assuming that the particle
possesses a X easy axis and XY easy plane , and the magnetic field $h$ is
applied along the hard axis (Z axis), the Hamiltonian operator of the AFM
particle has the form

\begin{equation}
\hat{H}=\stackunder{a=1,2}{\sum }\left( k_{\bot }\hat{S}_{{\small a}%
}^{z2}+k_{\shortparallel }\hat{S}_a^{y2}-\gamma h\hat{S}_a^z\right) +J\hat{S}%
_1\cdot \hat{S}_2
\end{equation}
where $k_{\bot },k_{\shortparallel }>0$ are the anisotropy constants, $J$ is
the exchange constant, $\gamma $ is the gyromagnetic ratio,and the spin
operators in two sublattices $\hat{S}_1$ and $\hat{S}_2$ obey the usual
commutation relation $\left[ \hat{S}_a^i,\hat{S}_b^j\right] =i\hbar \epsilon
_{ijk}\delta _{ab}\hat{S}_b^k\ \left( i,j,k=x,y,z;a,b=1,2\right) $. The
matrix element of the evolution operator in spin coherent-state
representation is

\begin{equation}
\langle N_f|e^{-2i\hat{H}T/\hbar }|N_i\rangle =\int \left[ \stackunder{k=1}{%
\stackrel{M-1}{\prod }}d\mu \left( \text{N}_k\right) \right] \left[ 
\stackunder{k=1}{\stackrel{M}{\prod }}\langle \text{N}_k|e^{-i\epsilon \hat{H%
}/\hbar }|\text{N}_{k-1}\rangle \right]
\end{equation}
Here we define $|$N$\rangle =|$n$_1\rangle |$n$_2\rangle $, $|$N$_M\rangle
=| $N$_f\rangle =|$n$_{1,f}\rangle |$n$_{2,f}\rangle ,$ $|$N$_o\rangle =|$N$%
_i\rangle =|$n$_{1,i}\rangle |$n$_{2,i}\rangle $, $t_f-t_i=2T$ and $\epsilon
=2T/M$. The spin coherent state is defined as

\begin{equation}
|\text{n}_a\rangle =e^{-i\theta _a\hat{C}}|S_a,S_a\rangle ,\left(
a=1,2\right)
\end{equation}
where n$_a=\left( \sin \theta _a\text{cos}\phi _a,\text{sin}\theta _a\text{%
sin}\phi _a,\text{cos}\theta _a\right) $ is the unit vector, \^{C}$_a=$sin$%
\phi _a\hat{S}_a^x-$cos$\phi _a\hat{S}_a^y$ and $|S_{a\text{,}}S_a\rangle $
is the reference spin eigenstate. The measure is defined by

\begin{equation}
d\mu \left( \text{N}_k\right) =\stackunder{a=1,2}{\prod }\frac{2S_a+1}{4\pi }%
d\text{n}_{a,k\text{ }},\qquad d\text{n}_{a,k}=\text{sin}\theta
_{a,k}d\theta _{a,k}d\phi _{a,k},
\end{equation}
In the large $S$ limit we obtain\cite{Nie}

\begin{equation}
\langle \text{N}_f|e^{-2i\hat{H}T/\hbar }|\text{N}_i\rangle =e^{-iS_0\left(
\phi _f-\phi _i\right) }\int \stackunder{a=1,2}{\prod }D[\theta _a]D[\phi
_a]\exp \left( \frac i\hbar \int_{t_i}^{t_f}Ldt\right)
\end{equation}
The Lagrangian is defined by $L=L_0+L_1$ with ~

\begin{equation}
L_0=\stackunder{a=1,2}{\sum }S_a\dot{\phi}_acos\theta _a~-JS_1S_2\left[ \sin
\theta _1\cos \theta _2\cos \left( \phi _1-\phi _2\right) +\cos \theta
_1\cos \theta _2\right]
\end{equation}

\begin{equation}
L_1=\stackunder{a=1,2}{\sum }\left( k_{\bot }S_a^2\cos ^2\theta
+k_{\shortparallel }S_a^2\sin ^2\theta _a\sin ^2\phi _a-\gamma hS_a\cos
\theta _a\right)
\end{equation}
where $S_0=S_1+S_2$ is total spin of two sublattices. Since $S_1$ and $S_2$
is almost antiparallel, we may replace $\theta _2$ and $\phi _2$ by $\theta
_2=\pi -\theta _1-\epsilon _{_\theta }$and $\phi _2=\pi +\phi _1+\epsilon
_{_\phi },$ where $\epsilon _{_\theta }$ and $\epsilon _{_\phi }$ denote
small fluctuations .Working out the fluctuation integrations over $\epsilon
_{_\theta }$ and $\epsilon _{_\phi }$ the transition amplitude Eq.(5)
reduces to

\begin{equation}
\langle \text{N}_f|e^{-2i\hat{H}T/\hbar }|\text{N}_i\rangle =e^{-iS_0\left(
\phi _f-\phi _i\right) }\int \stackunder{a=1,2}{\prod }D[\theta ]D[\phi
]\exp \left( \frac i\hbar \int_{t_i}^{t_f}\bar{L}dt\right)
\end{equation}

\begin{equation}
\bar{L}=\Omega \left[ \frac m\gamma \dot{\phi}\cos \theta +\frac{\chi _{\bot
}}\gamma H\dot{\phi}\text{sin}^2\theta +\frac \chi {2\gamma ^2}\left( \dot{%
\theta}^2+\dot{\phi}^2\sin ^2\theta \right) \right] -V\left( \theta ,\phi
\right) ,
\end{equation}
where $V\left( \theta ,\phi \right) =\left( \Omega K_{\bot }\cos ^2\theta
+K_{\shortparallel }\sin ^2\theta \sin ^2\phi -mh\cos \theta -\frac{\chi
_{\bot }}2h^2\sin ^2\theta \right) ,$and $\left( \theta _1,\phi _1\right) $
has been replaced by $\left( \theta ,\phi \right) $. $m=\gamma \hbar \left(
S_1-S_2\right) /\Omega $ with $\Omega $ being the volume of the AFM particle
and $\chi _{\bot }=\frac{\gamma ^2}J$. $K_{\bot }=2k_{\bot }S^2/\Omega $ and 
$K_{\shortparallel }=2k_{\shortparallel }S^2/\Omega $ (setting $S_1=S_2=S$
except in the term containing $S_1-S_2$) denote the transverse and the
longitudinal anisotropy constants, respectively.

We assume a very strong transverse anisotropy, $i.e.$ $K_{\bot }\gg $ $%
K_{\shortparallel }$ .In this case the N\'{e}el vector is forced to lie near
the XY plane. Replacing $\theta $ by $\frac \pi 2+$ $\eta $ where $\eta $
denotes the small fluctuation and carrying out the integral over $\eta $ we
obtain

\begin{equation}
\langle N_f|e^{-2\hat{H}\beta /\hbar }|N_i\rangle =\int D[\phi ]\exp \left(
-\frac 1\hbar \int_{\tau _i}^{\tau _f}L_{eff}d\tau \right)
\end{equation}
where

\begin{equation}
L_{eff}=\frac I2\left( \frac{d\phi }{d\tau }\right) ^2+i\Theta \frac{d\phi }{%
d\tau }+V\left( \phi \right)
\end{equation}
is the effective Euclidean Lagrangian. $\tau =it$ and $\beta =iT$ .$I=\Omega
\left( I_a+I_f\right) $ where $I_a=m^2/\left( 2\gamma ^2K_{\bot }\right) $
and $I_f=\chi _{\bot }/\gamma ^2$ are the effective FM and AFM moments of
inertia\cite{E.M.Chudnovsky}, respectively. $V\left( \phi \right) =\Omega
K_{\shortparallel }\sin ^2\phi $ is the effective potential and $\Theta
=\hbar S_0-I\gamma h$. The second term in Eq.(11) , $i.e.$ $i\Theta \frac{%
d\phi }{d\tau }$ which is the total time derivative has no effect on the
classical equation of motion, however it leads to a path dependent phase in
Euclidean action. The classical equation of motion is seen to be

\begin{equation}
\frac I2\left( \frac{d\phi }{d\tau }\right) ^2-V\left( \phi \right) =0
\end{equation}
$\phi =0$ and $\pi $ are two equilibrium orientation of the N\'{e}el vector.
The N\'{e}el vector may transit by tunneling through potential barriers from
one orientation ($\phi =0$) to another ($\phi =\pi $) along clockwise or
anticlockwise paths. The instanton solutions of Eq.(12) are then obtained as

\begin{equation}
\phi _c^{\pm }\left( \tau \right) =\pm 2\arctan \left( \text{e}^{\omega
_0\tau }\right)
\end{equation}
where $\omega _0=\sqrt{2K_{\shortparallel }\Omega /I}$ is the small
oscillation frequency of the N\'{e}el vector around its equilibrium
orientation. $\phi _c^{-}\left( \tau \right) $ and $\phi _c^{+}\left( \tau
\right) $ denote instanton solutions with clockwise and anticlockwise
windings respectively. The Euclidean actions evaluated along the instanton
trajectories are seen to be

\begin{equation}
S_E^{\pm }=\int L_{eff}d\tau =2I\omega _0\pm \Theta \pi
\end{equation}
The quantum phase interference of clockwise path ``$-$'' and anticlockwise
path ``+'' is seen to be (see Fig.1)

\begin{equation}
e^{-S_E^{+}}+e^{-S_E^{-}}\sim e^{-2I\omega _0/\hbar }\cos \left( \Lambda \pi
\right)
\end{equation}
where $\Lambda =\frac \Theta \hbar $ $=S_0+\frac h{h_c}$ with $h_c=\frac
\hbar {\gamma I}$ . Since the potential $V\left( \phi \right) $ is periodic
and can be regarded as a one-dimensional superlattice. Using the Bloch
theory the low-lying energy spectrum could be determined as\cite{J.Q.Liang.H}

\begin{equation}
E_0=\varepsilon _0-2\Delta \varepsilon _0\cos \pi \left( \Lambda +\xi \right)
\end{equation}
Where $\xi $ is Bloch wave vector which can be assumed to take either of the
two values 0 and 1\cite{J.Q.Liang.Y}. $\Delta \varepsilon _0=\frac{2\hbar
\omega _0}\pi $e$^{-2I\omega _0/\hbar }$ is the usual overlap integral or
simply the level shift induced by tunneling through any one of the barriers
. Thus the tunneling splitting is seen to be

\begin{equation}
\Delta \varepsilon =\left| 2\Delta \varepsilon _0\cos \pi \left( \Lambda
+1\right) -2\Delta \varepsilon _0\cos \pi \Lambda \right| =4\Delta
\varepsilon _0\left| \cos \pi \Lambda \right|
\end{equation}
which is a function of the external magnetic field like in the ferromagnetic
particle case \cite{A.Garg,W.Werns,S.P.Kou}.When $h=0$, Eq.(17) reduces to
the previous result\cite{E.M.C.in} where the tunneling splitting is quenched
when $S_0=$half-integer. With nonzero magnetic field the tunneling splitting
would be quenched whenever $\Lambda =n+\frac 12$ or $h=\left( S_0-n-\frac
12\right) \hbar /I\gamma $ where $n$ is an integer. Fig.2 shows the
oscillation of the tunneling splitting with respect to the external magnetic
field. This quenching is due to the quantum phase interference of two
tunneling paths of opposite windings. The period of oscillation is given by

\begin{equation}
\Delta h=\frac \hbar {I\gamma }.
\end{equation}

We have demonstrated a macroscopic quantum interference effect in the
tunneling of the magnetization of antiferromagnetic particles. Such
particles open thus a new avenue to test macroscopic quantum interference
effects. Experimental tests of our prediction could thus make an important
contribution to our understanding of the transition region between the
microscopic and the macroscopic world.

\begin{description}
\item  \textbf{Acknowledgements: }

This work was supported by the National Natural Science Foundation of China
under Grant Nos. 1967701 and 19775033. J.-Q.L and D.K.P also acknowledge
support by the Deutsche Forschungsgemeinschaft.
\end{description}

\textbf{Figure Caption:}

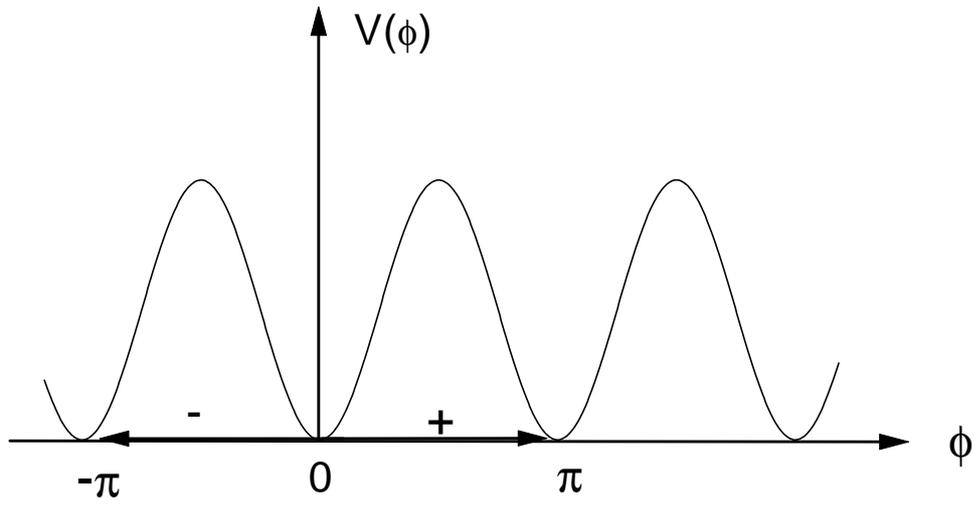
Fig.1: Quantum phase interference of two tunnel paths of opposite windings

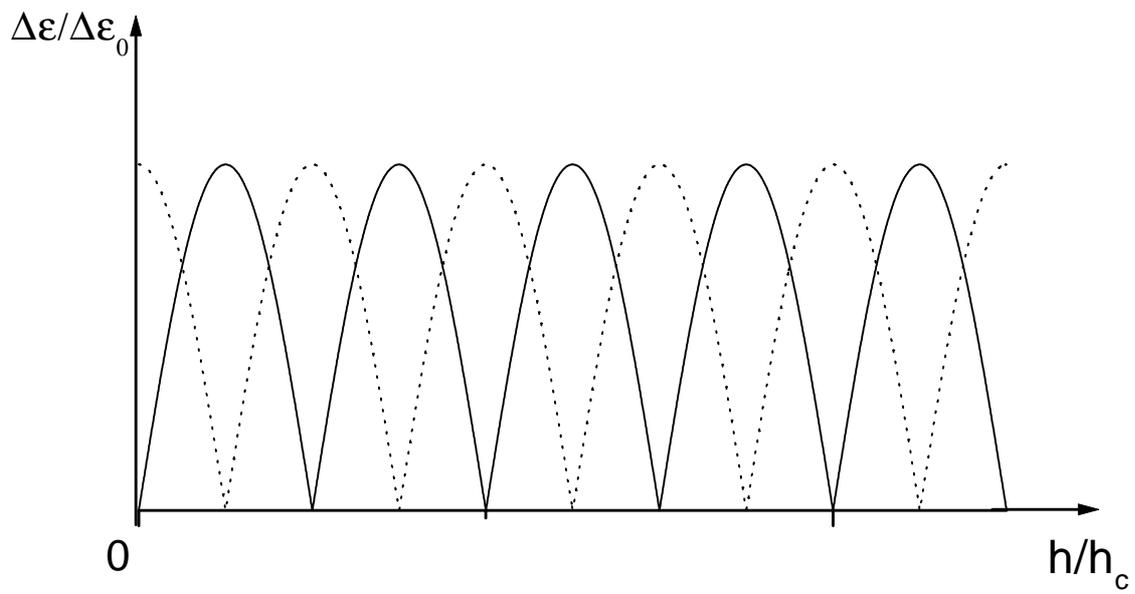
Fig.2: Oscillation of tunneling splitting as a function of the external
field with the solid line for $S_0=$half-integer and the dotted line for $%
S_0=\func{integer}$

\end{document}